\documentclass[prd,twocolumn,showpacs,preprintnumbers,amsmath,amssymb,superscriptaddress,floatfix,nofootinbib]{revtex4}

\usepackage{graphicx,color,dcolumn,booktabs,bm}
\usepackage{longtable,lscape}
\usepackage{txfonts}
\usepackage{overpic}
\usepackage{epsfig}
\usepackage{amssymb}
\usepackage{rotating}
\usepackage{epstopdf}
\usepackage{indentfirst}
\usepackage{feynmf}   
\usepackage{slashed}  
\usepackage{cases}
\usepackage{color}
\usepackage{multirow}
\usepackage{graphicx,color,dcolumn,booktabs,bm}
\usepackage{cases}
\usepackage{array}

\newcommand{\Slash}[1]{\ooalign{\hfil/\hfil\crcr$#1$}}

\begin{document}

\title{Strong decay mode $J/\psi p$ of hidden charm pentaquark states $P_c^+(4380)$ and $P_c^+(4450)$ in $\Sigma_c \bar{D}^*$ molecular scenario}

\author{Qi-Fang L\"{u}} \email{lvqifang@ihep.ac.cn}
\affiliation{Institute of High Energy Physics, Chinese Academy of Sciences, Beijing 100049, China}
\author{Yu-Bing Dong} \email{dongyb@ihep.ac.cn}
\affiliation{Institute of High Energy Physics, Chinese Academy of Sciences, Beijing 100049, China}
\affiliation{Theoretical Physics Center for Science Facilities (TPCSF), CAS, Beijing 100049, China}

\begin{abstract}

We study the strong decay channels $P_c^+(4380) \to J/\psi p$ and $P_c^+(4450) \to J/\psi p$ under $\Sigma_c \bar{D}^*$ molecular state ansatz. With various spin-parity assignments, the partial decay widths of $J/\psi p$ final state are calculated. Our results show that all the $P$ wave $\Sigma_c {\bar D}^*$ assignments are excluded, while $S$ wave $\Sigma_c {\bar D}^*$ pictures for $P_c(4380)$ and $P_c(4450)$ are both allowed by present experimental data. The $J^P=3/2^-$ $\Sigma_c^* \bar{D}$ and $\Sigma_c^* \bar{D}^*$ molecules are also discussed in the heavy quark limit, and find that the $\Sigma_c^* \bar{D}$ system for $P_c(4380)$ is possible. More experimental information on spin-parities and partial decay widths and theoretical investigations on other decay modes are needed to clarify the nature of the two $P_c$ states.

\end{abstract}
\maketitle

\section{Introduction}{\label{introduction}}

In 2015, two hidden charm pentaquark resonances $P_c^+(4380)$ and $P_c^+(4450)$ were reported by LHCb Collaboration in the $J/\psi p$ invariant mass spectrum in the $\Lambda_b^0 \to J/\psi K^- p$ process~\cite{Aaij:2015tga}. Their masses and widths are
\begin{eqnarray}
(M,\Gamma)_{P_c^+(4380)} = (4380\pm8\pm29,205\pm18\pm86)~\rm{MeV},
\end{eqnarray}
\begin{eqnarray}
(M,\Gamma)_{P_c^+(4450)} = (4449.8\pm1.7\pm2.5,39\pm5\pm19)~\rm{MeV}.
\end{eqnarray}
The spin-parities of the $P_c$ states are not determined yet, and some possible spin-parity values $(3/2^-, 5/2^+)$, $(3/2^+, 5/2^-)$, and $(5/2^+, 3/2^-)$ for $[P_c(4380), P_c(4450)]$ are favored by the experimental analyses. Given the strong decay mode $J/\psi p$, the $P_c(4380)$ and $P_c(4450)$ states are believed to have five intrinsic quark content $uudc\bar{c}$, and their isospins are $I = 1/2$.

In the following paper, LHCb Collaboration reported the branching fraction of the decay $\Lambda_b^0 \to J/\psi K^- p$ and branching ratios ${\cal B}(\Lambda_b^0 \to P_c^+ K^-){\cal B}(P_c^+ \to J/\psi p)$~\cite{Aaij:2015fea},
\begin{eqnarray}
&&{\cal B}(\Lambda_b^0 \to J/\psi K^- p) \nonumber \\
&&= 3.04\pm0.04\pm0.06\pm0.33^{+0.43}_{-0.27} \times 10^{-4},
\end{eqnarray}
\begin{eqnarray}
&&{\cal B}(\Lambda_b^0 \to P_c^+(4380) K^-){\cal B}(P_c^+(4380) \to J/\psi p) \nonumber \\
&&= 2.56\pm0.22\pm1.28^{+0.46}_{-0.36} \times 10^{-5},
\end{eqnarray}
\begin{eqnarray}
&&{\cal B}(\Lambda_b^0 \to P_c^+(4450) K^-){\cal B}(P_c^+(4450) \to J/\psi p) \nonumber \\
&&= 1.25\pm0.15\pm0.33^{+0.22}_{-0.18} \times 10^{-5}.
\end{eqnarray}
However, the available experimental data are insufficient to determine the values of ${\cal B}(P_c^+ \to J/\psi p)$.

In addition to the conventional mesons ($q\bar{q}$) and baryons ($qqq$), the exotic states,
such as tetraquark, pentaquark, and dibaryons, are also allowed by color confinement. In the past decades, lots of $XYZ$ states have been observed, which can be hardly described as conventional mesons in traditional quark models~\cite{Agashe:2014kda,Liu:2013waa,Chen:2016qju}. Most of them, especially the $Z_c^+$ and $Z_b^+$ states, should have four quark content. A structure of $d^*(2380)$ has been reported in double-pion fusion reactions and their partial wave analyses~\cite{Bashkanov:2008ih,Adlarson:2014pxj}, which may have six quark content. The observation of two pentaquark states $P_c(4380)$ and $P_c(4450)$ provides a unique perspective to the nature of exotic states.

Before the discovery of $P_c$ states, there have already been some theoretical studies on the existence of hidden charm pentaquark states with meson-baryon and meson-meson interactions~\cite{Wu:2010jy,Wu:2010vk,Yang:2011wz,Wang:2011rga,Garcia-Recio:2013gaa,Xiao:2013yca,Uchino:2015uha}. Some hidden charm states are predicted, which may correspond the $P_c$ states or their partners. Moreover, the productions of these states are also discussed within the framework of effective Lagrangian approach~\cite{Wu:2012wta,Huang:2013mua,Garzon:2015zva,Wang:2015xwa}.

Inspired by the observation of $P_c$ states, many interesting theoretical works have been presented. According to the diquark picture, the two $P_c$ states are interpreted as compact pentaquarks, where the five quarks are tightly bound and interact with each other via color magnetic interaction directly~\cite{Maiani:2015vwa,Wang:2015ava,Wang:2015epa,Lebed:2015tna,Lebed:2015dca,Anisovich:2015zqa,Zhu:2015bba}. An alternative explanation is that the $P_c$ resonances are molecular states composed of one meson and one baryon, which are loosely bounded and interact through color Van der Waals interaction and boson exchange interaction~\cite{Karliner:2015ina,Chen:2015loa,Roca:2015dva,He:2015cea,Chen:2015moa,Meissner:2015mza,Huang:2015uda,Eides:2015dtr,Chen:2016otp}. Furthermore, the non-resonant interpretations of these two structures also exist, which suggest that the $P_c$ states are not genuine states, but kinematical effects due to re-scattering processes~\cite{Guo:2015umn,Liu:2015fea,Mikhasenko:2015vca,Liu:2016dli}. Detailed discussions of various assignments and properties can be found in Ref.~\cite{Chen:2016qju,Burns:2015dwa}.

The production mechanism of these two $P_c$ are also investigated by some works. In analogy to $\Lambda_b^0 \to J/\psi K^- p$, the $P_c$ states and their partners can be produced in weak decay processes of bottom baryons~\cite{Cheng:2015cca,Li:2015gta,Feijoo:2015cca,Hsiao:2015nna,Chen:2015sxa,Wang:2015pcn,Oset:2016lyh}. Since the anomalous triangle singularity kinematical effects do not exist in the two body final states reactions, the photon and pion induced productions are suggested to clarify the nature of $P_c$ states~\cite{Wang:2015jsa,Kubarovsky:2015aaa,Karliner:2015voa,Lu:2015fva}. Other production processes, such as heavy ion collisions~\cite{Wang:2016vxa}, proton-nucleus collisions~\cite{Schmidt:2016cmd}, and $\Upsilon(1S) \to p \bar{p} J/\psi$ reaction~\cite{Guo:2015umn}, are also proposed.

Besides the mass spectrum and production mechanism, the decay behaviors of $P_c$ states can also provide beneficial information of their structures. However, there are only a few studies on decay properties. Burns discussed possible decay patterns of $P_c$ states under various scenarios according to quantum number conservations~\cite{Burns:2015dwa}. Eides, Petrov and Polyakov calculated the $J/\psi$ partial decay width of $P_c(4450)$ as a bound state of charmonium $\psi(2S)$ and nucleon. Wang, Li, Liu and Zhu performed a comprehensive study with the heavy quark symmetry and spin rearrangement scheme, where the $\Sigma_c^* \bar{D}$, $\Sigma_c \bar{D}^*$, and $\Sigma_c^* \bar{D}^*$ molecular candidates are considered~\cite{Wang:2015qlf}. Independent of dynamical mechanism, several ratios of partial widths of some decay channels are estimated. Particularly, Wu, Molina, Oset, and Zou gave a pioneer investigation on decay behaviors of the predicted hidden charm states~\cite{Wu:2010jy}.

It should be mentioned that the masses of $\Sigma_c \bar{D}^*$ systems with different partial waves have been extensively studied in the literature. Some works suggested that $\Sigma_c \bar{D}^*$ molecule stands for the $P_c(4380)$ state~\cite{Chen:2015loa,Chen:2015moa,Chen:2016otp}, while others support it as $P_c(4450)$ resonance~\cite{Karliner:2015ina,He:2015cea}. In the present work, we will investigate the strong decay mode $J/\psi p$ of $P_c$ states under $\Sigma_c \bar{D}^*$ molecule assignments with spin-parity $J^P=3/2^\pm, 5/2^\pm$. Within the framework of effective Lagrangians approach, the technique for evaluating composite hadron systems have been widely used for various molecular states, where the compositeness condition, corresponding to $Z=0$ has been employed~\cite{Dong:2008,Dong:2009a,Dong:2009tg,Dong:2013kta,Dong:2014zka,Gutsche:2014zda,Chen:2015igx}. Moreover, the $J^P=3/2^-$ $\Sigma_c^* \bar{D}$ and $\Sigma_c^* \bar{D}^*$ molecules are also discussed in the heavy quark limit. These explorations can give the predicted partial decay widths of $P_c$ states, which will help us to distinguish between different interpretations to be tested by future experiments.

This paper is organized as follows.  The basic formalism and ingredients of our approach are displayed in Sec.~\ref{formalism}. The predicted partial decay widths and discussions are presented in Sec.~\ref{results}. Finally, we give a short summary in the last section.

\section{Formalism and Ingredients}{\label{formalism}}

In this section, we present the formalism for the investigation of $P_c$ states with various spin-parity assignments in the $\Sigma_c \bar{D}^*$ molecular pictures. The simplest effective Lagrangian describing the $P_c\Sigma_c \bar{D}^*$ couplings can be expressed as
\begin{eqnarray}
{\cal L}_{P_c\Sigma_c \bar{D}^*}^{{3/2}^\pm}(x)&=&-i g_{P_c\Sigma_c \bar{D}^*}\int d^4y \Phi(y^2) \bar {\Sigma}_c(x+\omega_{\bar{D}^*} y) \Gamma^{(\pm)} \nonumber \\
&& \bar{D}^*_{\mu}(x-\omega_{\Sigma_c} y) P_c^{\mu}(x) + \rm{H.c}.,
\end{eqnarray}
\begin{eqnarray}
{\cal L}_{P_c\Sigma_c \bar{D}^*}^{{5/2}^\pm}(x)&=&g_{P_c\Sigma_c \bar{D}^*}\int d^4y \Phi(y^2) \bar {\Sigma}_c(x+\omega_{\bar{D}^*} y) \Gamma^{(\mp)} \nonumber \\&& \partial_{\mu} \bar{D}^*_{\nu}(x-\omega_{\Sigma_c} y) P_c^{\mu\nu}(x) + \rm{H.c}.,
\end{eqnarray}
where $\omega_{\bar{D}^*} = m_{\bar{D}^*}/(m_{\Sigma_c}+m_{\bar{D}^*})$ and $\omega_{\Sigma_c} = m_{\Sigma_c}/(m_{\Sigma_c}+m_{\bar{D}^*})$. The vertex $\Gamma^{(\pm)}$ matrix is defined as $\Gamma^+ = \gamma_5$ and $\Gamma^- = 1$. The effective correlation function $\Phi(y^2)$ is chosen to describe the distribution of $\Sigma_c$ and $\bar{D}^*$ in $P_c$ states. To render the Feynman diagrams ultraviolet finite, the Fourier transform of correlation function $\Phi(y^2)$ should vanishes fast in the ultraviolet region of the Euclidean space. We adopt the commonly used forms in hadronic molecular decays and the Fourier transform can be written as~\cite{Dong:2008,Dong:2009a,Dong:2009tg,Dong:2013kta,Dong:2014zka,Gutsche:2014zda,Chen:2015igx}
\begin{eqnarray}
\tilde\Phi(p^2_E /\Lambda^2) \doteq {\rm{exp}}(-p^2_E /\Lambda^2),
\end{eqnarray}
where $p_E$ is the Euclidean Jacobi momentum and $\Lambda$ is a free size parameter characterizing the distribution of the two components in the molecule. This parameter is about 1 GeV and varies in different systems.

The coupling constant $g_{P_c\Sigma_c \bar{D}^*}$ can be determined by the compositeness condition~\cite{Weinberg:1962hj,Salam:1962ap}. This condition requires that the renormalization constant of the hadronic molecular wave function is equal to zero,
\begin{eqnarray}
Z_{P_c} = 1 - \frac{\partial\Sigma_{P_c}^{T(\pm)}(p)}{\partial \Slash p} {\bigg |}_{\Slash p=m_{P_c}} \equiv 0,
\end{eqnarray}
where the $\Sigma_{P_c}^{T(\pm)}(p)$ is the transverse part of the mass operator $\Sigma_{P_c}^{\mu\nu(\pm)}(p)$ with the $J^P=3/2^{\pm}$ assumption and $\Sigma_{P_c}^{\mu\nu\alpha\beta(\pm)}(p)$ with the $J^P=5/2^{\pm}$ assumption of the molecule $P_c$ states, The relation between the transverse part and the corresponding mass operator can be defined as
\begin{eqnarray}
\Sigma_{P_c}^{\mu\nu(\pm)}(p) &=& g_{\perp}^{\mu\nu} \Sigma_{P_c}^{T(\pm)}(p) + \frac{p^\mu p^\nu}{p^2} \Sigma_{P_c}^{L(\pm)}(p),
\end{eqnarray}
\begin{eqnarray}
\Sigma_{P_c}^{\mu\nu\alpha\beta(\pm)}(p) &=& \frac{1}{2}(g^{\mu\alpha}_{\perp} g^{\nu\beta}_{\perp}+g^{\mu\beta}_{\perp} g^{\nu\alpha}_{\perp})\Sigma_{P_c}^{T(\pm)}(p)+ \cdot \cdot \cdot,
\end{eqnarray}
with $g^{\mu\nu}_{\perp}=g^{\mu\nu}-p^\mu p^\nu /p^2$. The $\cdot \cdot \cdot$ denote terms that do not contribute to the mass renormalization.

\begin{figure}[htbp]
\includegraphics[scale=0.6]{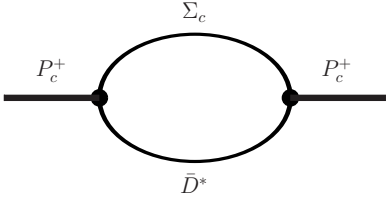}
\vspace{0.0cm} \caption{The mass operator of $P_c^+$ states.}
\label{mass}
\end{figure}

The diagram describing the mass operator of the $P_c$ states is presented in Fig.~\ref{mass}. From the effective Lagrangian, we can obtain the concrete forms of $\Sigma_{P_c}^{\mu\nu(\pm)}(p)$ and $\Sigma_{P_c}^{\mu\nu\alpha\beta(\pm)}(p)$,
\begin{eqnarray}
\Sigma_{P_c}^{\mu\nu(\pm)}(p) &=& \pm g_{P_c\Sigma_c \bar{D}^*}^2 \int \frac{d^4q}{(2\pi)^4 i} \tilde \Phi^2(q-\omega_{\bar{D}^*}p) \Gamma^{(\pm)} \frac{1}{\Slash q - m_{\Sigma_c}}  \nonumber \\&& \times  \Gamma^{(\pm)} \frac{-g^{\mu\nu}+(p-q)^{\mu}(p-q)^{\nu}/m_{\bar{D}^*}^2}{(p-q)^2-m_{\bar{D}^*}^2},
\end{eqnarray}
\begin{eqnarray}
\Sigma_{P_c}^{\mu\nu\alpha\beta(\pm)}(p) &=& \pm g_{P_c\Sigma_c \bar{D}^*}^2 \int \frac{d^4q}{(2\pi)^4 i} \tilde \Phi^2(q-\omega_{\bar{D}^*}p) \Gamma^{(\mp)} \nonumber \\&& \times \frac{1}{\Slash q - m_{\Sigma_c}} \Gamma^{(\mp)} (p-q)^\alpha(p-q)^\mu \nonumber \\&& \times \frac{-g^{\beta\nu}+(p-q)^{\beta}(p-q)^{\nu}/m_{\bar{D}^*}^2}{(p-q)^2-m_{\bar{D}^*}^2} ,
\end{eqnarray}
where $q$ is the four momentum of $\Sigma_c$ baryon.

\begin{figure*}[!htp]
\includegraphics[scale=0.6]{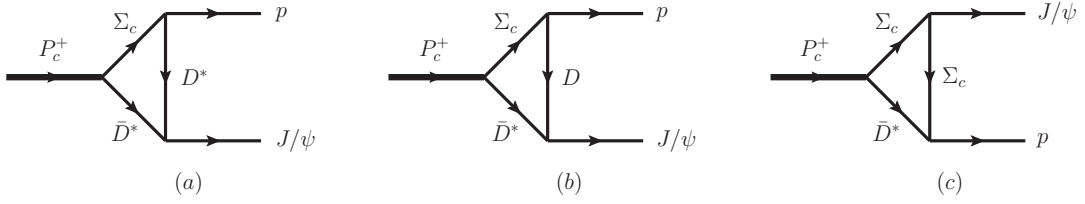}
\vspace{0.0cm} \caption{Feynman diagrams for the $P_c^+ \to J/\psi p$ decay process.}
\label{feyn}
\end{figure*}

The hadronic decay $P_c^+ \to J/\psi p$ occurs by exchanging $D^*$, $D$, and $\Sigma_c$ hadrons as shown in Fig.~\ref{feyn}. To estimate these triangle loop diagrams, the effective Lagrangian of relevant interaction vertices are also needed~\cite{Garzon:2015zva,Chen:2015igx,Casalbuoni:1996pg},
\begin{eqnarray}
{\cal L}_{\psi D^{(*)} D^{(*)}} &=& -i g_{\psi D D} \psi_{\mu}(\partial^{\mu} D D^\dag - D \partial^{\mu} D^\dag)\nonumber
\\&& + g_{\psi D^* D} \varepsilon^{\mu\nu\alpha\beta} \partial_{\mu} \psi_{\nu} (D^*_{\alpha} \overleftrightarrow \partial_{\beta} D^\dag - D \overleftrightarrow \partial_{\beta} D^{*\dag}_{\alpha})\nonumber
\\&& + i g_{\psi D^* D^*} \psi^{\mu} (D^*_{\nu} \partial^{\nu} D^{*\dag}_{\mu} - \partial^{\nu} D^*_{\mu} D^{*\dag}_{\nu} \nonumber
\\&& - D^*_{\nu} \overleftrightarrow \partial_{\mu} D^{*\nu\dag}),
\end{eqnarray}
\begin{eqnarray}
{\cal L}_{\Sigma_c N D^*} &=& g_{\Sigma_c N D^*} \bar{N} \gamma_{\mu} {\vec \tau} \cdot {\vec \Sigma_c} D^{*\mu}+ \rm{H.c}.,
\end{eqnarray}
\begin{eqnarray}
{\cal L}_{\Sigma_c N D} &=& -i g_{\Sigma_c N D} \bar{N} \gamma_5 {\vec \tau} \cdot {\vec \Sigma_c} D+ \rm{H.c}.,
\end{eqnarray}
\begin{eqnarray}
{\cal L}_{\Sigma_c \Sigma_c \psi} &=& g_{\Sigma_c \Sigma_c \psi} {\bar{\Sigma}}_c \gamma_{\mu} \Sigma_c \psi^{\mu}.
\end{eqnarray}

In the heavy quark limit, the coupling constants of charmonium and charmed mesons should satisfy~\cite{Chen:2015igx,Casalbuoni:1996pg}
\begin{eqnarray}
g_{\psi D D} &=& 2 g_2 \sqrt{m_\psi} m_D,\nonumber
\\ g_{\psi D^* D} &=& 2 g_2 \sqrt{m_D m_{D^*} /m_\psi},\nonumber
\\ g_{\psi D^* D^*} &=& 2 g_2 \sqrt{m_\psi} m_{D^*},
\end{eqnarray}
where $g_2=\sqrt{m_\psi}/(2m_Df_\psi)$ is a gauge coupling and $f_\psi$ is the decay constant of a charmonium $\psi$. For the $J/\psi$, one can obtain $f_{J/\psi} = 426~\rm{MeV}$ with the leptonic partial decay width. Based on $SU(4)$ invariant Lagrangians and flavor symmetry, we can determined the coupling constants $g_{\Sigma_c N D^*}=g_{\Sigma_c \Sigma_c \psi}=3.0$~\cite{Dong:2009tg}, and $g_{\Sigma_c N D} = 2.69$~\cite{Garzon:2015zva}.

With the above Lagrangians, the amplitudes for $P_c^+(p_0) \to [\Sigma_c(q_1) \bar{D}^*(q_2)] \to p(p_1)J/\psi(p_2)$ can be written straightforwardly
\begin{eqnarray}
{\cal M}_a^{3/2^{\pm}} &=& -i g_{P_c\Sigma_c \bar{D}^*} g_{\Sigma_c N D^*} g_{\psi D^* D^*} \int \frac{d^4q}{(2\pi)^4} \nonumber
\\&& \times \tilde \Phi(\omega_{\bar D^*}q_1 - \omega_{\Sigma_c} q_2) \epsilon^{*\nu}_{\psi} \bar u_N \gamma_\alpha \frac{1}{\Slash q_1 - m_{\Sigma_c}} \nonumber \\&& \times \Gamma^{\prime(\pm)} u^{\mu}_{P_c}  [q_\rho g_{\nu\lambda} - q_{2\lambda} g_{\rho\nu} + (q_{2\nu} - q_\nu) g_{\rho\lambda}]
\nonumber \\&& \times \frac{-g^\rho_\mu + q_{2\mu}q_2^\rho/m_{D^*}^2}{q_2^2-m_{D^*}^2} \frac{-g^{\lambda \alpha} + q^\lambda q^\alpha/m_{D^*}^2}{q^2-m_{D^*}^2},
\end{eqnarray}
\begin{eqnarray}
{\cal M}_b^{3/2^{\pm}} &=& g_{P_c\Sigma_c \bar{D}^*} g_{\Sigma_c N D} g_{\psi D^* D} \int \frac{d^4q}{(2\pi)^4} \nonumber
\\&& \times \tilde \Phi(\omega_{\bar D^*}q_1 - \omega_{\Sigma_c} q_2) \epsilon^{*\nu}_{\psi} \bar u_N \gamma_5 \frac{1}{\Slash q_1 - m_{\Sigma_c}} \nonumber \\&& \times \Gamma^{\prime(\pm)} u^{\mu}_{P_c} \epsilon_{\rho \lambda \alpha \nu} p_2^{\alpha} (q_\lambda-q_{2\lambda})
\nonumber \\&& \times \frac{-g^\rho_\mu + q_{2\mu}q_2^\rho/m_{D^*}^2}{q_2^2-m_{D^*}^2} \frac{1}{q^2-m_D^2},
\end{eqnarray}
\begin{eqnarray}
{\cal M}_c^{3/2^{\pm}} &=& i g_{P_c\Sigma_c \bar{D}^*} g_{\Sigma_c N D^*} g_{\Sigma_c \Sigma_c \psi} \int \frac{d^4q}{(2\pi)^4} \nonumber
\\&& \times \tilde \Phi(\omega_{\bar D^*}q_1 - \omega_{\Sigma_c} q_2) \epsilon^{*\nu}_{\psi} \bar u_N \gamma_{\rho} \frac{1}{\Slash q - m_{\Sigma_c}} \gamma_{\nu} \nonumber \\&& \times  \frac{1}{\Slash q_1 - m_{\Sigma_c}}  \Gamma^{\prime(\pm)} u^{\mu}_{P_c}
 \frac{-g^\rho_\mu + q_{2\mu}q_2^\rho/m_{D^*}^2}{q_2^2-m_{D^*}^2},
\end{eqnarray}
\begin{eqnarray}
{\cal M}_a^{5/2^{\pm}} &=&  g_{P_c\Sigma_c \bar{D}^*} g_{\Sigma_c N D^*} g_{\psi D^* D^*} \int \frac{d^4q}{(2\pi)^4} \nonumber
\\&& \times \tilde \Phi(\omega_{\bar D^*}q_1 - \omega_{\Sigma_c} q_2) \epsilon^{*\nu}_{\psi} \bar u_N \gamma_\alpha \frac{1}{\Slash q_1 - m_{\Sigma_c}} \nonumber \\&& \times \Gamma^{\prime(\mp)} u^{\mu\beta}_{P_c} q_{2\beta} [q_\rho g_{\nu\lambda} - q_{2\lambda} g_{\rho\nu} + (q_{2\nu} - q_\nu) g_{\rho\lambda}]
\nonumber \\&& \times \frac{-g^\rho_\mu + q_{2\mu}q_2^\rho/m_{D^*}^2}{q_2^2-m_{D^*}^2} \frac{-g^{\lambda \alpha} + q^\lambda q^\alpha/m_{D^*}^2}{q^2-m_{D^*}^2},
\end{eqnarray}
\begin{eqnarray}
{\cal M}_b^{5/2^{\pm}} &=& i g_{P_c\Sigma_c \bar{D}^*} g_{\Sigma_c N D} g_{\psi D^* D} \int \frac{d^4q}{(2\pi)^4} \nonumber
\\&& \times \tilde \Phi(\omega_{\bar D^*}q_1 - \omega_{\Sigma_c} q_2) \epsilon^{*\nu}_{\psi} \bar u_N \gamma_5 \frac{1}{\Slash q_1 - m_{\Sigma_c}} \nonumber \\&& \times \Gamma^{\prime(\mp)} u^{\mu\beta}_{P_c} \epsilon_{\rho \lambda \alpha \nu} q_{2\beta} p_2^{\alpha} (q_\lambda-q_{2\lambda})
\nonumber \\&& \times \frac{-g^\rho_\mu + q_{2\mu}q_2^\rho/m_{D^*}^2}{q_2^2-m_{D^*}^2} \frac{1}{q^2-m_D^2},
\end{eqnarray}
\begin{eqnarray}
{\cal M}_c^{5/2^{\pm}} &=& - g_{P_c\Sigma_c \bar{D}^*} g_{\Sigma_c N D^*} g_{\Sigma_c \Sigma_c \psi} \int \frac{d^4q}{(2\pi)^4} \nonumber
\\&& \times \tilde \Phi(\omega_{\bar D^*}q_1 - \omega_{\Sigma_c} q_2) \epsilon^{*\nu}_{\psi} \bar u_N \gamma_{\rho} \frac{1}{\Slash q - m_{\Sigma_c}} \gamma_{\nu} \nonumber \\&& \times  \frac{1}{\Slash q_1 - m_{\Sigma_c}}  \Gamma^{\prime(\mp)} u^{\mu\beta}_{P_c} q_{2\beta}
 \frac{-g^\rho_\mu + q_{2\mu}q_2^\rho/m_{D^*}^2}{q_2^2-m_{D^*}^2},
\end{eqnarray}
where $q$ is the four momentum of the exchanged hadrons. The vertex $\Gamma^{\prime(\pm)}$ matrix is defined as $\Gamma^{\prime+} = i \gamma_5$ and $\Gamma^{\prime-} = 1$. Then the total amplitudes of the $P_c^+ \to J/\psi p$ process are,
\begin{eqnarray}
{\cal M} &=& {\cal M}_a+{\cal M}_b+{\cal M}_c.
\end{eqnarray}

The explicit expression of the $P_c^+$ states in molecule pictures can be expressed as~\cite{Wu:2010vk}
\begin{eqnarray}
P_c^+=\sqrt{\frac13}\Sigma_c^+\bar{D}^{*0}+\sqrt{\frac23}\Sigma_c^{++}D^{*-}.\label{Pc}
\end{eqnarray}
The isospin spin symmetry requires the coupling of the $\Sigma_c^{++}pD^{*+}$ vertex is $\sqrt{2}$ times larger than the one of the $\Sigma_c^{+}pD^{*0}$. Then the corresponding partial decay widths can be estimated by
\begin{eqnarray}
\Gamma(P_c^+ \to J/\psi p) &=& \frac{1}{2J+1}\frac{3}{8\pi}\frac{|{\vec p}_1|}{m_{P_c}^2}\overline{|{\cal M}|^2},
\end{eqnarray}
where $J$ is the total angular momentum of initial $P_c$ states, the overline indicates the sum over the polarization vectors of hadrons, and ${\cal M}$ stands for the contribution from $\Sigma_c^+\bar{D}^{*0}$. The factor 3 comes from the two components of Eq.~(\ref{Pc}) where the mixing is included. Certainly, one can absorb this factor into the amplitudes~\cite{Wu:2010vk}.

\section{Results and Discussions}{\label{results}}

\begin{figure}[!htbp]
\includegraphics[width=8.5cm, height=8cm]{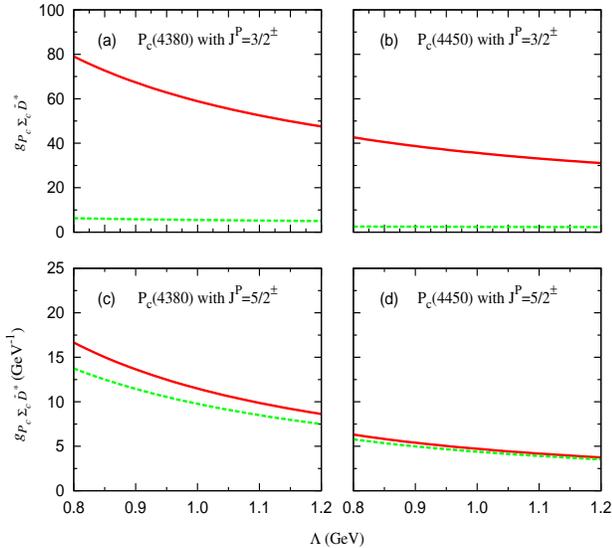}
\vspace{-0.8cm} \caption{The coupling constants of $P_c$ states with different $J^P$ assignments depending on the parameter $\Lambda$. The red solid line stands for $P=+$, and green dashed line corresponds $P=-$ cases.}
\label{coup}
\end{figure}

The coupling constants $g_{P_c \Sigma_c {\bar D}^*}$ of two $P_c$ states with different spin-parity assignments are estimated in the $\Sigma_c {\bar D}^*$ hadronic molecular pictures from the compositeness condition. The typical value of $\Lambda$ is about 1 GeV, and the $\Lambda$ dependence of these coupling constants are shown in Fig.~\ref{coup}. When $\Lambda$ varies from 0.8 to 1.2 GeV, the coupling $g_{P_c \Sigma_c {\bar D}^*}$ decreases. It is known that for a $S$ wave loosely bound state, the effective coupling strength of the bound state to its components is insensitive to the inner structure and can be determined model-independently~\cite{Weinberg:1962hj,Baru:2003qq}. In our calculation, the coupling constants are almost independent with the parameter $\Lambda$ for the $J^P = 3/2^-$ cases, where the $\Sigma_c {\bar D}^*$ system is in $S$ wave. The $\Lambda$ dependence indicates the realistic systems in our calculations are not so loosely bound, and in the case of more deeply bound state the dependence of the coupling on $\Lambda$ becomes stronger. Comparing the couplings of $P_c(4380)$ and $P_c(4450)$ states with the same $J^P$ assumptions, it is also shown that for a certain $\Lambda$, the coupling constant decreases as the binding energy decreases. The values of $g_{P_c \Sigma_c {\bar D}^*}$ with $\Lambda = 1~ {\rm GeV}$ are listed in Table.~\ref{tab1}.

\begin{table}[!htbp]
\begin{center}
\caption{ \label{tab1} Coupling constants of $P_c \Sigma_c {\bar D}^*$ with different $J^P$ assignments with $\Lambda = 1~ {\rm GeV}$.}
\small
\begin{tabular*}{8.5cm}{@{\extracolsep{\fill}}*{5}{p{1.55cm}<{\centering}}}
\hline\hline
  State               & $3/2^+$       & $3/2^-$   & $5/2^+$    &$5/2^-$ \\
  & & &($\rm{GeV^{-1}}$) &($\rm{GeV^{-1}}$)\\   \hline
  $P_c(4380)$         & 58.91         &5.48       & 11.49       &9.77  \\
  $P_c(4450)$         & 35.66         &2.38       & 4.71        &4.39  \\
\hline\hline
\end{tabular*}
\end{center}
\end{table}

\begin{figure}[!htbp]
\includegraphics[scale=0.86]{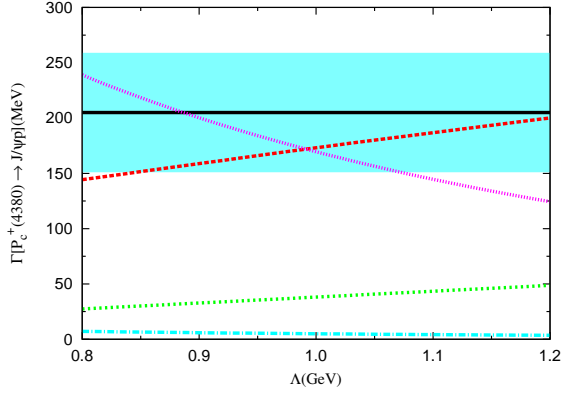}
\vspace{0.0cm} \caption{The partial decay widths of the $P_c^+(4380) \to J/\psi p$ with different $J^P$ assignments depending on the parameter $\Lambda$. The red dashed, green dotted, pink short dotted, and blue dot-dashed lines stand for $J^P = 3/2^+,~3/2^-,~5/2^+~\rm{and}~5/2^-$ cases, respectively. The black solid line and blue error band correspond to the total width observed by experiment.}
\label{4380}
\end{figure}

\begin{figure}[!htbp]
\includegraphics[scale=0.86]{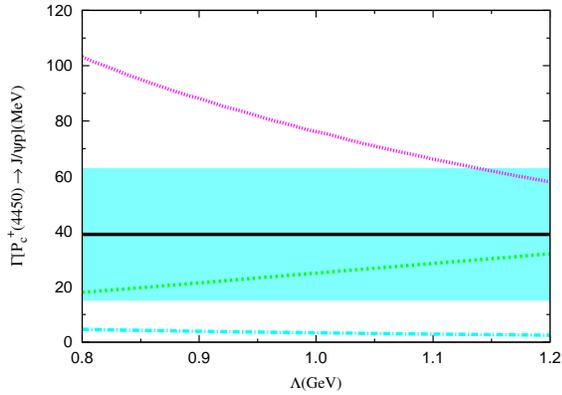}
\vspace{0.0cm} \caption{The same as Fig.~\ref{4380}, but for the $P_c^+(4450) \to J/\psi p$. The partial decay width of $J^P = 3/2^+$ case is much larger than the experimental data and neglected here.}
\label{4450}
\end{figure}

The partial decay widths of $P_c^+ \to J/\psi p$ are presented in Fig.~\ref{4380} and Fig.~\ref{4450}, and their values with $\Lambda = 1~ {\rm GeV}$ are shown in Table.~\ref{tab2}. In the discussed $\Lambda$ range, the partial decay widths increase with $\Lambda$ for the $J^P = 3/2^\pm$ assignments, while decrease for the $J^P = 5/2^\pm$ assignments. For the $P_c(4380)$ state, the predicted partial decay widths in the four cases are all smaller than the experimental total width within the limit of error-bar. The $J^P = 5/2^-$ assignment is not favored by LHCb experiment~\cite{Aaij:2015tga}, and the $\Sigma_c {\bar D}^*$ molecule should be in $D$ wave in this case. Hence, the $J^P = 5/2^-$ assumption for $P_c(4380)$ is excluded, while others are consistent with the present data. For the $P_c(4450)$ state, since $J^P = 3/2^+$ assignment is not favored by LHCb collaboration~\cite{Aaij:2015tga} and the calculated partial decay width is much larger than the experimental total width, this case can be completely excluded as well. Also, the $J^P=5/2^+$ case is disfavored due to the large predicted width, and the $D$ wave $\Sigma_c {\bar D}^*$ molecule with $J^P = 5/2^-$ can be hardly formed through boson exchange interaction and color Van der Waals interaction. Hence, only the $S$ wave $\Sigma_c {\bar D}^*$ molecule is left for the $P_c(4450)$ state in our calculation.

\begin{table}[!htbp]
\begin{center}
\caption{ \label{tab2} Partial decay widths of $P_c^+ \to J/\psi p$ with different $J^P$ assignments with $\Lambda = 1~ {\rm GeV}$. The unit is in MeV. }
\small
\centering
\begin{tabular*}{8.5cm}{@{\extracolsep{\fill}}*{5}{p{1.4cm}<{\centering}}}
\hline\hline
  State               & $3/2^+$       & $3/2^-$   & $5/2^+$     &$5/2^-$   \\\hline
  $P_c(4380)$         & 173.12         &38.12       & 169.51      &4.96   \\
  $P_c(4450)$         & 369.82         &25.00       & 76.15       &3.39  \\
\hline\hline
\end{tabular*}
\end{center}
\end{table}

\begin{figure}[!htbp]
\includegraphics[width=8.5cm, height=8cm]{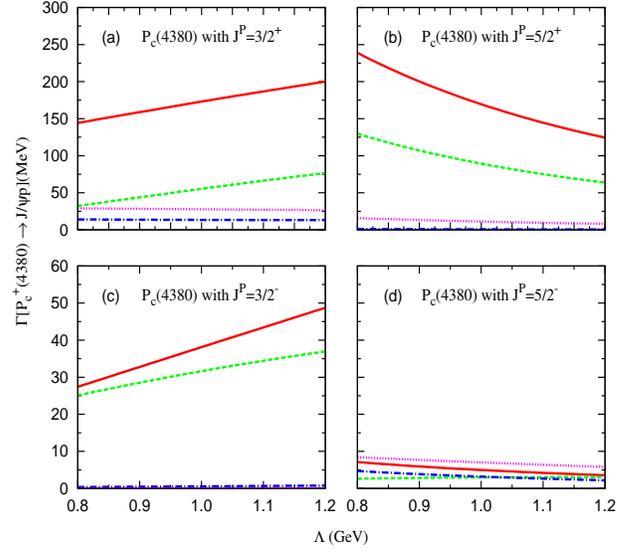}
\vspace{0.0cm} \caption{The individual contributions of the $D^*$, $D$, and $\Sigma_c$ exchanges for the $P_c^+(4380) \to J/\psi p$ with different $J^P$ assignments. The red solid, green dotted, blue dot-dashed, and pink short dotted lines stand for the total, $D^*$, $D$, and $\Sigma_c$ contributions, respectively.}
\label{4380abc}
\end{figure}

The individual contributions of the $D^*$, $D$, and $\Sigma_c$ exchanges for the $P_c^+(4380) \to J/\psi p$ (for an example) with different $J^P$ assignments are presented in Fig.~\ref{4380abc} . Since the relative signs of the three Feynman diagrams are well
defined, the obtained total decay widths are the square of their coherent sum. It is found that for the $S$ and $P$ wave cases, the diagram of $D^*$ exchange plays dominated role, while $D$ and $\Sigma_c$ exchanges give minor contributions. However, the interferences among them are still sizable. Even for the $J^P = 3/2^-$ case, the $\mid{\cal M}_b + {\cal M}_c \mid$ is about $1/10 \sim 1/5$ of the $\mid{\cal M}_a\mid$, which contributes significantly constructive interferences. The behaviors of each contributions for $P_c(4450)$ are similar to the ones of the $P_c(4380)$, since only the binding energy is changed.

Besides the $\Sigma_c {\bar D}^*$ molecular picture, the $\Sigma_c^* \bar D$ and $\Sigma_c^* {\bar D}^*$ scenarios are also proposed in the literature. We do not calculate, in this work,  the partial decay widths of $P_c$ states as $\Sigma_c^* \bar D$ and $\Sigma_c^* {\bar D}^*$ molecules due to the extremely complicated and lengthy procedures of loop integrals including higher spin states. In the heavy quark limit, the $S$-wave $D$ and $D^*$ mesons can be categorized into a doublet as well as the heavy baryons $\Sigma_c$ and $\Sigma_c^*$. With the heavy quark symmetry and spin rearrangement scheme, there is an interesting work studying the ratios of partial decay widths in different molecular scenarios~\cite{Wang:2015qlf}. The model independent result shows that for the three $S$-wave $\Sigma_c {\bar D}^*$, $\Sigma_c^* \bar D$, and $\Sigma_c^* {\bar D}^*$ molecules with $J^P = 3/2^-$, the ratios of their $J/\psi p$ decay widths satisfy $\Gamma[(\Sigma_c {\bar D}^*)]:\Gamma[(\Sigma_c^* \bar D)]:\Gamma[(\Sigma_c^* {\bar D}^*)] = 1.0:2.7:5.4$. Simply employing those ratios and the results listed in Tab.~\ref{tab2}, we can estimate the $J/\psi p$ decay widths of the two $P_c$ states in $\Sigma_c^* \bar D$ and $\Sigma_c^* {\bar D}^*$ molecular scenarios with $J^P = 3/2^-$,
\begin{eqnarray}
&& \Gamma_{P_c^+(4380) \to J/\psi p}^{3/2^-} [(\Sigma_c^* \bar D)] = 102.92~\rm{MeV},\nonumber
\\&& \Gamma_{P_c^+(4380) \to J/\psi p}^{3/2^-} [(\Sigma_c^* {\bar D}^*)] = 205.85~\rm{MeV},\nonumber
\\&& \Gamma_{P_c^+(4450) \to J/\psi p}^{3/2^-} [(\Sigma_c^* \bar D)] = 67.50~\rm{MeV},\nonumber
\\&& \Gamma_{P_c^+(4450) \to J/\psi p}^{3/2^-} [(\Sigma_c^* {\bar D}^*)] = 135.00~\rm{MeV}.
\label{limit}
\end{eqnarray}
The much larger partial decay width $\Gamma_{P_c^+(4450) \to J/\psi p}^{3/2^-} [(\Sigma_c^* {\bar D}^*)]$ excludes the possibility of $P_c(4450)$ as a $S$-wave $J^P = 3/2^-$ $\Sigma_c^* {\bar D}^*$ molecule. Also, the $P_c(4450)$ as the $\Sigma_c^* \bar D$ system is not favored due to its higher mass over the threshold and a slightly large partial decay width. The above discussion shows that if $P_c(4450)$ state has the spin-parity $J^P = 3/2^-$, only the $\Sigma_c {\bar D}^*$ system of the three molecular scenarios is allowed. This result is consistent with the interpretations of Ref.~\cite{Karliner:2015ina}, in which the $P_c(4450)$ is a $J^P = 3/2^-$ $\Sigma_c {\bar D}^*$ resonance and $P_c(4380)$ may be not a genuine state.

It is worthy mentioned that the $J/\psi p$ mode of the two $P_c$ states may not be the dominated decay channel. The situation occurs for the charged charmonium-like and bottomonium-like states. Under the assumption that $Z_c(3885)$ and $Z_c(3900)$ are the same state, the partial decay width of $D {\bar D}^*$ channel is larger than $J/\psi \pi$ mode by a factor about 6~\cite{Ablikim:2013xfr}. The open-bottom decay modes of $Z_b(10610)$ and $Z_b(10650)$ are also dominated~ \cite{Agashe:2014kda}. In Refs.~\cite{Chen:2016qju,Karliner:2016via}, an argument also suggests that the $P_c$ states as $\Sigma_c^{(*)} \bar{D}^{(*)}$ molecules should mainly decay to open charm final states rather than $J/\psi p$. From Tab.~\ref{tab2} and Eq.~(\ref{limit}), three partial decay widths $\Gamma_{P_c^+(4380) \to J/\psi p}^{3/2^+} [(\Sigma_c {\bar D}^*)]$, $\Gamma_{P_c^+(4380) \to J/\psi p}^{5/2^+} [(\Sigma_c {\bar D}^*)]$, and $\Gamma_{P_c^+(4380) \to J/\psi p}^{3/2^-} [(\Sigma_c^* {\bar D}^*)]$ seem too large, and the branching ratios are over $80\%$, which indicate the $J/\psi p$ decay modes dominate. These results stand in sharp contrast to the above discussions. Hence, we believe that the $P_c(4380)$ as these assignments are disfavored.

It should be stressed that a very recent work calculates the relative decay ratio of the $\bar D^* \Lambda_c$ and $J/\psi p$ states for the $P_c(4380)$ in both $\Sigma_c \bar D^*$ and $\Sigma_c^* \bar D$ pictures~\cite{Shen:2016tzq}. The ratio is about one for the $\Sigma_c \bar D^*$ case, but more than 20 for the $\Sigma_c^* \bar D$ assignment. The authors suggest that the $\Sigma_c^* \bar D$ ansatz is more reasonable than the $\Sigma_c \bar D^*$ one in order to interpret the broad width of $P_c(4380)$. In our present work, both $\Sigma_c \bar D^*$ and $\Sigma_c^* \bar D$ pictures are allowed. If we combine the ratios of Ref.~\cite{Shen:2016tzq} and the partial decay widths of $J/\psi p$ in the present work, the $\bar D^* \Lambda_c$ widths can be obtained. The partial decay width of $\bar D^* \Lambda_c$ is about 38 MeV for the $\Sigma_c \bar D^*$ case, while it is more than 2000 MeV for the $\Sigma_c^* \bar D$ case. This excludes the $\Sigma_c^* \bar D$ ansatz for $P_c(4380)$. These different conclusions may arise from different perspectives and the model dependent results, and we prefer to believe that the both $\Sigma_c \bar D^*$ and $\Sigma_c^* \bar D$ scenarios possible.

To sum up, the $S$-wave $\Sigma_c {\bar D}^*$ assignments for $P_c(4380)$ and $P_c(4450)$ are both allowed by our calculations, while all the $P$-wave $\Sigma_c {\bar D}^*$ assumptions are excluded. Moreover, for the $J^P=3/2^-$ $\Sigma_c^* \bar D$ and $\Sigma_c^* {\bar D}^*$ scenarios, only $\Sigma_c^* \bar D$ for $P_c(4380)$ case is possible. It should be noted that these results do not contradict with the fact that the two $P_c$ states have different spins and opposite parities, since other cases such as $P$-wave $\Sigma_c^* {\bar D}^*$ assumption are also possible and not discussed here. Theoretical investigations on other decay modes and further experimental information on spin-parities and partial decay widths will be helpful to distinguish these possible assignments of the $P_c$ states.

\section{Summary}{\label{summary}}

In this work, we estimate the partial decay widths of strong decay modes $J/\psi p$ of hidden charm pentaquark states $P_c(4380)$ and $P_c(4450)$ in $\Sigma_c \bar{D}^*$ molecular scenarios. With various spin-parity assignments, the partial decay widths of $P_c$ states are significantly different. All the $P$ wave $\Sigma_c {\bar D}^*$ assignments are excluded, while $S$ wave $\Sigma_c {\bar D}^*$ pictures for $P_c(4380)$ and $P_c(4450)$ are both possible. The $J^P=3/2^-$ $\Sigma_c^* \bar{D}$ and $\Sigma_c^* \bar{D}^*$ molecular pictures are also discussed in heavy quark limit, and only $\Sigma_c^* \bar D$ for $P_c(4380)$ is allowed. More theoretical investigations on other decay behaviors and experimental information on spin-parities and partial decay widths are needed to clarify the nature of two $P_c$ states.

\bigskip
\noindent
\begin{center}
{\bf ACKNOWLEDGEMENTS}\\
\end{center}

This project is supported by the National Natural Science Foundation of China under Grants No.~10975146, and No.~11475192. The fund provided by the Sino-German CRC 110 ``Symmetries and the Emergence of Structure in QCD" project (NSFC Grant No. 11261130311) is also appreciated.

\end{document}